\documentclass[11pt,a4paper]{article}
\usepackage[hyperref]{acl2020}
\usepackage{times}
\usepackage{latexsym}
\usepackage{graphicx}
\usepackage{amsmath}
\usepackage{hyperref}

\usepackage{tabularx}
\usepackage{caption}
\usepackage{float}
\usepackage{multirow}
\usepackage{subfigure}
\usepackage{microtype}

\aclfinalcopy

\title{Learning-Based Compression for Machines}

\author{Kartik Gupta\textsuperscript{$\dagger$}, Kimberley Faria, Vikas Mehta\\
    University of Massachusetts Amherst\\
    \textsuperscript{$\dagger$}\texttt{kgupta@umass.edu}}

\date{}

\begin{document}
\maketitle

\begin{abstract}
While learning-based compression techniques for images have outperformed traditional methods, they have not been widely adopted in machine learning pipelines. This is largely due to lack of standardization and lack of  retention of salient features needed for such tasks. Decompression of images have taken a back seat in recent years while the focus has shifted to an image's utility in performing machine learning based analysis on top of them. Thus the demand for compression pipelines that incorporate such features from images has become ever-present.   The methods outlined in the report build on the recent work done on learning based image compression techniques to incorporate downstream tasks in them. We propose various methods of fine-tuning and enhancing different parts of pre-trained compression encoding pipeline and present the results of our investigation regarding the performance of vision tasks using compression based pipelines. 
\end{abstract}

\section{Introduction}
\subsection{Task Description}

Image and video compression is a crucial component of every content delivery pipeline, as it decreases storage costs and network transfer times. While historically compressed content was created uniquely with human consumption in mind, this has changed in recent years as content analytics ML-models become more ubiquitous. This machine-centric consumption includes, just to name a few, elderly monitoring in healthcare situations or traffic control and vehicle monitoring in transportation scenarios. In the era of machine learning, we can expect most of image and videos to be processed for ML downstream tasks in addition to be viewed by humans.

Nevertheless, current image compression pipelines come from several decades of development with human consumption in mind: image compression algorithms are designed to minimize visual distortion and make the content appealing for the human eye. Moreover, current ML processing pipelines usually require decoding the image before it can be analyzed (i.e., they operate in the pixel space rather than in the compressed space), which is an expensive and latency consuming operation. On the other hand, when ML processing is expected, compression should take this aspect into account to maximize both the quality of the final output from a human perspective and, at the same time, maximize the efficiency of downstream ML tasks. The goal of this project is to explore the performance of these early learning-based compression approaches that are designed both for human and machine consumption. Hence we will be exploring how information from compressed images can be used for machine learning tasks for example classification. We would also be exploring the formulation of an end to end pipeline consisting of a compression module(encoder) and a classification module for images by training these two deep learning based modules for specific downstream tasks. This is an active area of research and we will be exploring how we can improve performance for the same. 

\subsection{Motivation and Limitations of Existing Work}

Recent standardization effort led by JPEG\cite{jpegai} are starting to tackle this issue and create learning-based image compression pipelines where the learned compressed representation can be used directly into downstream ML tasks, without the need for decoding. While learned image compression methods like \citet{https://doi.org/10.48550/arxiv.1802.01436} achieve better compression performance, decoding images for downstream tasks is computationally expensive. Additionally, these models are trained with objective functions based on human judgement (PSNR, SSIM).  \citet{https://doi.org/10.48550/arxiv.2104.10065} adopted the framework proposed by \citet{jpegai} to the task of texture and material recognition and trained a truncated Resnet \cite{https://doi.org/10.48550/arxiv.1803.06131} to operate on compressed latent codes for the downstream task. Their results on the compressed domain space are satisfactory, however they did not match up to the results obtained when directly operating on the decoded or original images. Other work and ideas involve adding a pre-processing step to highlight/emphasize certain features that are useful for solving the downstream task. \citet{https://doi.org/10.48550/arxiv.2206.05650} added a preprocessing before encoding the original image. However, it still requires that the downstream task be performed after decoding, which is would be computationally expensive as the number of images increase. \citet{9956532} utilized a gate module to select suitable channel and remove the redundancy of the compressed domain representation for machine vision tasks to reduce the bit rates required for encoding. In addition, knowledge distillation is introduced to improve the accuracy of machine vision tasks. Other work has also focused on certain feature selection to improve downstream task performance \cite{9707070}. 

\subsection{Proposed Solution}

Our proposed solution uses the latent representation of image as a an input for and the compression pipeline including the encoder and downstream tasks are jointly trained for a downstream task. We also show the effectiveness of the model for satellite images. The results will be evaluated based on the accuracy of these specific tasks as well as the compression rate i.e. the number of bits required to represent an image. We utilize the proposed model presented in \citet{https://doi.org/10.48550/arxiv.2104.10065} and show effectiveness of the model for other tasks aswell like satellite images. Then we jointly train the encoder and downstream model for the task of texture recognition and compare the results.  
\begin{itemize}
    \item For compressing texture recognition images use the bmshj2018\_hyperprior model which is a Neural Network based compression model
    \item Use the compressed images and labeled outputs to train a cResNet-39 model to classify these compressed images
    \item This includes data augmentation as mentioned in \citet{https://doi.org/10.48550/arxiv.2104.10065}
    \item Jointly train encoder and downstream model for the machine learning task. 
\end{itemize}

The aforementioned joint training requires passing loss functions from downstream cResNet-39 model towards the hyperprior network for fine tuning which outputs the compressed representation and standard deviation feature maps for each image and losses are passed through these layers for training the encoder aswell as the downstream model. This process is done for 3 different settings of bits per pixel output by the encoder hence for each representational setting we have a different model which is fine tuned accordingly. A comparitive study about them and the downstream model with a frozen encoder gives us an informed insight in this pipeline. 
This provides the foundation of the framework proposed by JPEG AI and allows us to easily make modifications/improvements over this.

 Experiments in the domain indicate end to end training of compression and classification models in conjunction achieve better performance due to the compression module learning to focus on features that will help the downstream model.

\section{Experiments}

\subsection{Downstream Tasks and Datasets}
Our proposed machine vision tasks involve different classifications for images from several datasets. Separately, for each task we will use the following dataset.
\begin{itemize}
    \item For texture detection, we have used Materials in Context (MINC) Database, specifically, its subset MINC-2500\cite{bell2015material}
    \item For image classification we will use the \cite{resisc} dataset which consists of two sets of satellite images namely RESISC-45 and RSCNN-7 each containing different terrains classified based on satellite images. 
\end{itemize} 
All these datasets are publicly available.

\subsection{Baseline}
We follow these broad baselines for our step-by-step experiments. This includes the task of texture recognition using frozen encoder and a trained downstream model.Training a model which solves a particular task (classification/recognition) which takes as input the compressed-domain representation with 3 settings of quality/compression rate: bpp (bits per pixel) in $[1, 4, 8]$. Specifically we attempt to reproduce the results for texture recognition on the compressed domain representation \cite{https://doi.org/10.48550/arxiv.2104.10065}.

\begin{table*}[ht!]
\centering
\begin{tabular}{|c || c | c | c|} 
 \hline
 \textbf{Top-1 ACC} & \textbf{HyperMS-SSIM-1} & \textbf{HyperMS-SSIM-4} & \textbf{HyperMS-SSIM-8} \\ [0.5ex] 
 \hline
 \citet{https://doi.org/10.48550/arxiv.2104.10065} & 72.59 & 73.03 & 76.56 \\ 
 \hline
 Ours & 13.88 & 43.58 & 61.36 \\
 \hline\hline
 \textbf{Top-5 ACC} & \textbf{HyperMS-SSIM-1} & \textbf{HyperMS-SSIM-4} & \textbf{HyperMS-SSIM-8} \\ [0.5ex] 
 \hline
 \citet{https://doi.org/10.48550/arxiv.2104.10065} & 94.66 & 94.78 & 95.84 \\ 
 \hline
 Ours & 42.82 & 76.14 & 89.81 \\ [1ex] 
 \hline
\end{tabular}
\caption{Comparison of Baseline results in the compressed domain representation setting}
\label{table:1}
\end{table*}

\begin{figure*}
\centering
  \includegraphics[scale=0.7]{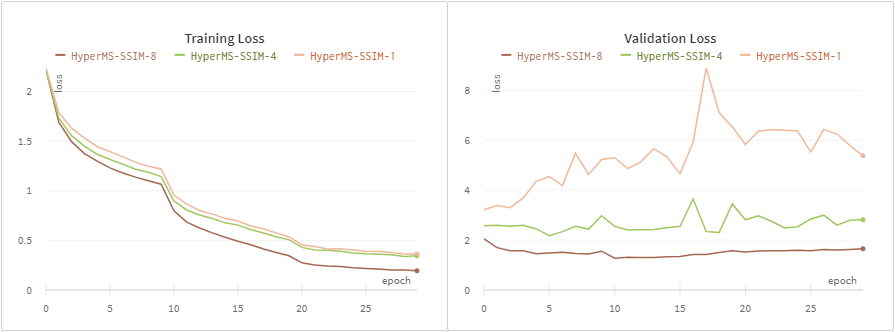}
  \caption{Training and Validation Loss Baseline}
  \label{fig1.1}
\end{figure*} 

\begin{figure*}
\centering
  \includegraphics[scale=0.7]{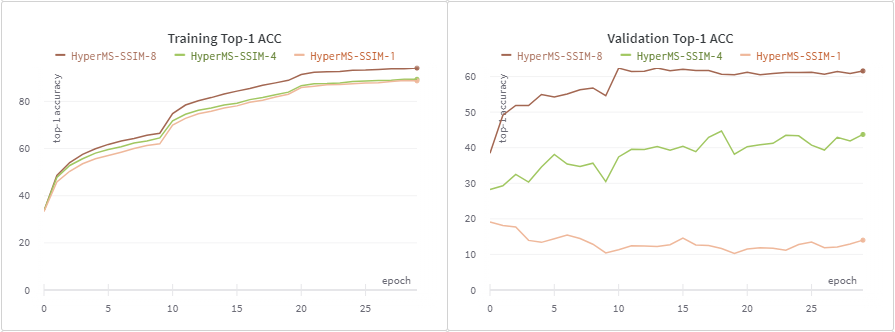}
  \caption{Top 1 Accuracy for Training and Validation sets Baseline}
  \label{fig1.2}
\end{figure*} 

\begin{figure*}
\centering
  \includegraphics[scale=0.2, width=\textwidth]{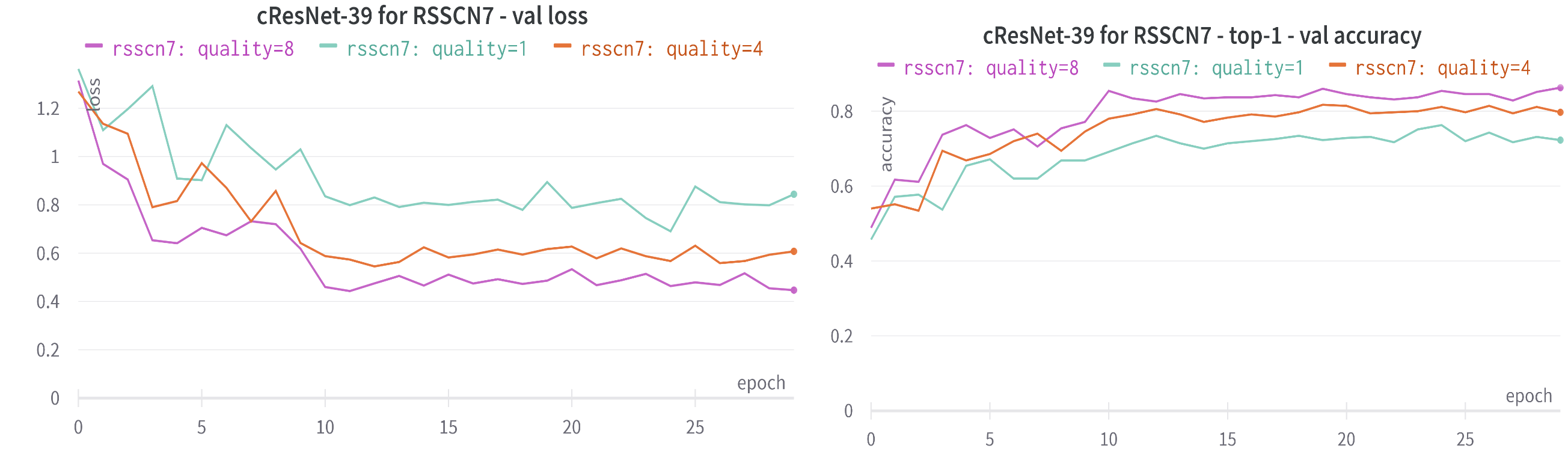}
  \caption{cResNet-39 Validation results for RSCNN7 dataset}
  \label{fig2}
\end{figure*}

\begin{figure*}
\centering
  \includegraphics[scale=0.4, width=\textwidth]{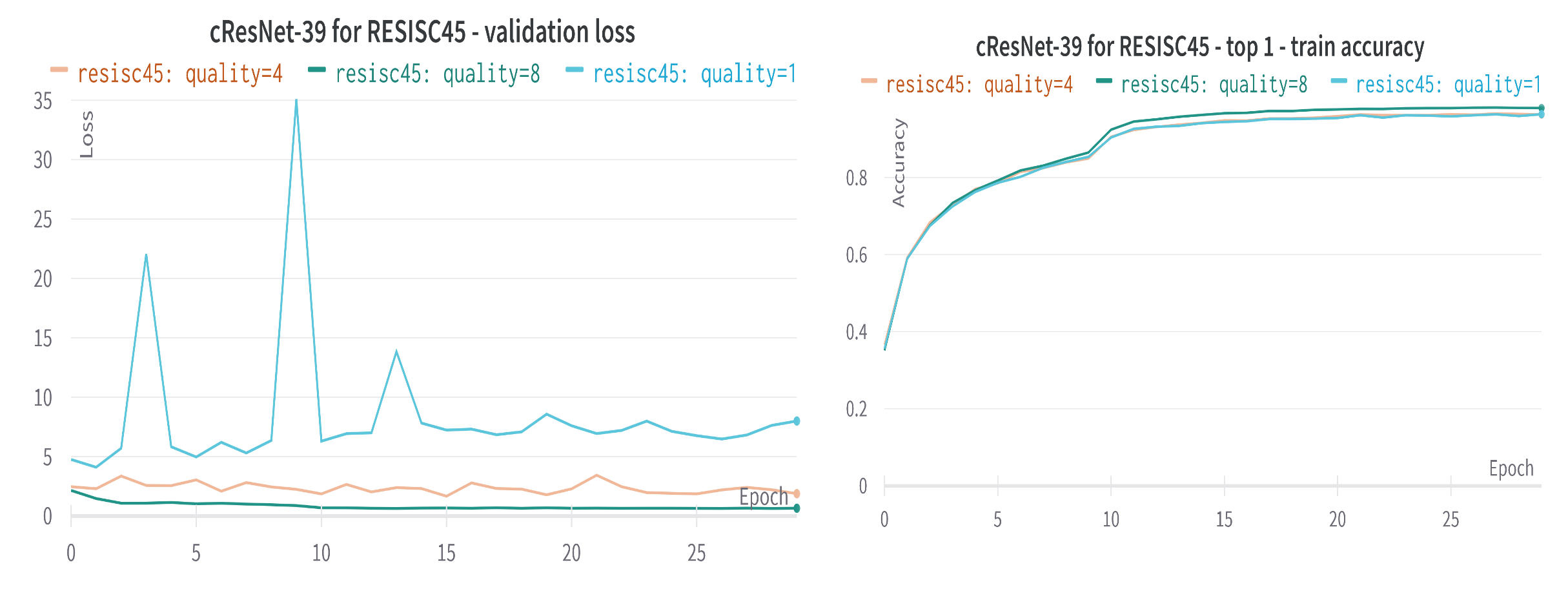}
  \caption{cResNet-39 Validation results for RESISC-45 dataset}
  \label{fig3}
\end{figure*}

\begin{figure*}
\centering
  \includegraphics[scale=0.2, width=\textwidth]{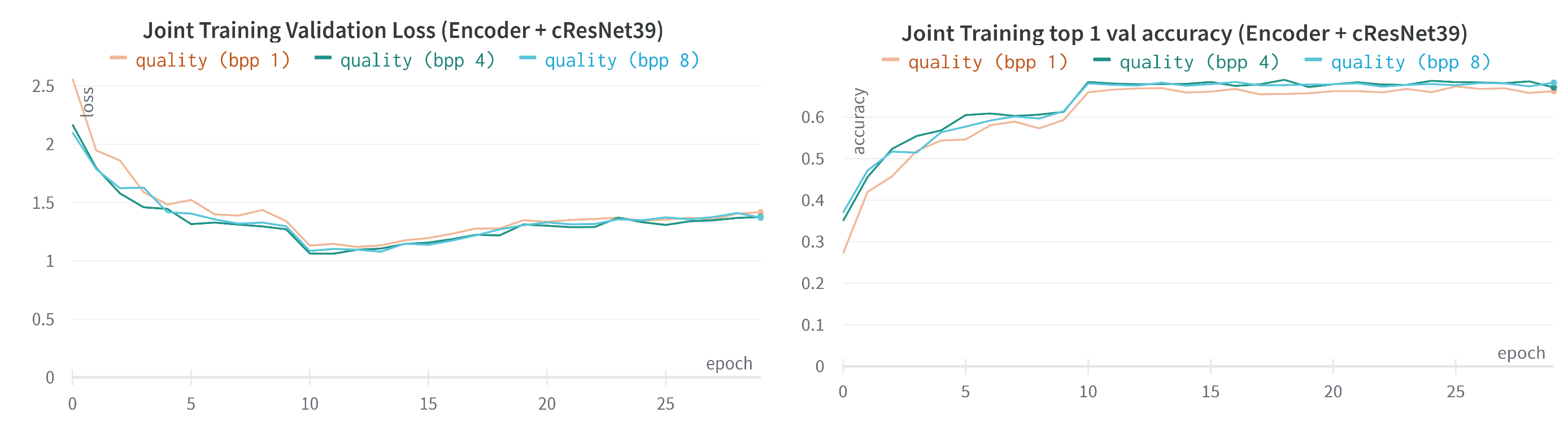}
  \caption{Validation results for joint training on texture recognition}
  \label{fig4}
\end{figure*}

\begin{table}[ht!]
\centering
\begin{tabular}{|c | c|} 
 \hline
 \textbf{Layer} & \textbf{cResNet-39} \\ [0.5ex] 
 \hline
 conv1 & None  \\ 
 \hline
 conv\_2x & 
 $\begin{bmatrix} 
 1 \times 1, 32 \\ 
 3 \times 3, 32 \\ 
 1 \times 1, 128 \\ 
 \end{bmatrix}$ 
  $\begin{bmatrix} 
 1 \times 1, 32 \\ 
 3 \times 3, 32 \\ 
 1 \times 1, 128 \\ 
 \end{bmatrix}$ \\ 
 \hline
 conv\_3x & 
 $\begin{bmatrix} 
 1 \times 1, 128 \\ 
 3 \times 3, 128 \\ 
 1 \times 1, 512 \\ 
 \end{bmatrix} \times 4$ \\ 
 \hline
 conv\_4x & 
 $\begin{bmatrix} 
 1 \times 1, 256 \\ 
 3 \times 3, 256 \\ 
 1 \times 1, 1024 \\ 
 \end{bmatrix} \times 6$ \\
 \hline
 conv\_5x & 
 $\begin{bmatrix} 
 1 \times 1, 512 \\ 
 3 \times 3, 512 \\ 
 1 \times 1, 2048 \\ 
 \end{bmatrix} \times 3$ \\
 \hline
  & average pool, 23-d fc, softmax \\ 
 \hline
\end{tabular}
\caption{cResNet-39 model architecture}
\label{table:2}
\end{table}

\begin{figure}
\centering
  \includegraphics[scale=0.47]{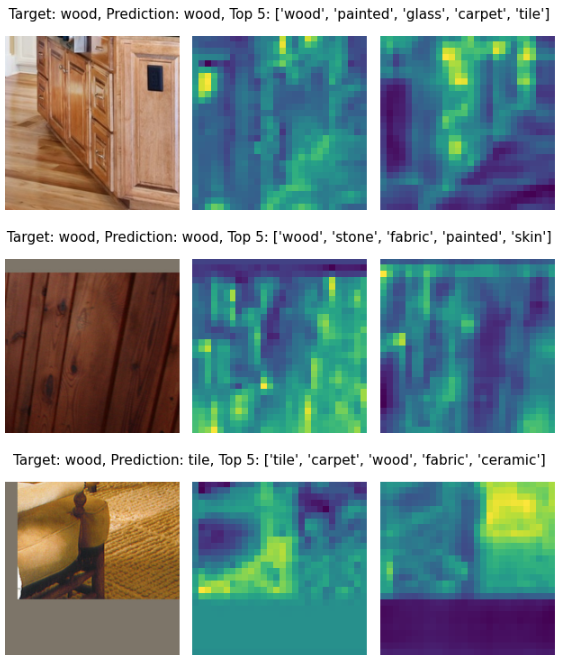}
  \caption{Predictions on the validation set. From left to right, we have the original raw image, compressed latent representation $\hat{y}$, standard deviations $\hat{\sigma}$ }
  \label{fig1.4}
\end{figure}

\begin{table*}[htbp]
\centering
\renewcommand{\arraystretch}{1.5} 
\renewcommand{\tabcolsep}{15pt}
\captionsetup{position=bottom}
\begin{tabular}{|c|*{4}{c|}}
\hline
\multirow{2}{*}{\textbf{Quality (BPP)}} & \multicolumn{2}{c|}{\textbf{RSCNN7}} & \multicolumn{2}{c|}{\textbf{RESISC45}} \\
\cline{2-5}
 & \textbf{Val Top-1 Acc} & \textbf{Val Top-5 Acc} & \textbf{Val Top-1 Acc} & \textbf{Val Top-5 Acc} \\
\hline
1 &  72.29& 99.4& 10.71& 26.22 \\
\hline
4 &  79.77& 100& 57.09& 87.36 \\
\hline
8 &  86.29& 100& 81.91*& 97.47
 \\
\hline
\end{tabular}
\caption{Results for cResNet-39 for satellite images}
\label{table:3}
\end{table*}

\begin{table*}[htbp]
\centering
\renewcommand{\arraystretch}{1.5} 
\renewcommand{\tabcolsep}{15pt}
\captionsetup{position=bottom}
\begin{tabular}{|c|*{4}{c|}}
\hline
\multirow{2}{*}{\textbf{Quality (BPP)}} & \multicolumn{2}{c|}{\textbf{cResNet-39}} & \multicolumn{2}{c|}{\textbf{Joint Training}} \\
\cline{2-5}
 & \textbf{Val Top-1 Acc} & \textbf{Val Top-5 Acc} & \textbf{Val Top-1 Acc} & \textbf{Val Top-5 Acc} \\
\hline
1 &  13.88&42.82&66.26&91.34\\
\hline
4 &  43.58&76.14&67.13&91.93\\
\hline
8 &  61.36&89.81&68.28&92.17\\
\hline
\end{tabular}
\caption{Results for joint training and baseline}
\label{table:4}
\end{table*}

\subsection{Implementation}

We have implemented the first baseline mentioned above. The code base to recreate the experiments in \citet{https://doi.org/10.48550/arxiv.2104.10065} was not readily available.
We attempted to recreate the setup in \citet{https://doi.org/10.48550/arxiv.2104.10065} by first creating compressed image outputs for the MINC-2500 dataset, i.e. we encode the images using the learning-based compression model proposed \citet{https://doi.org/10.48550/arxiv.1802.01436} and save these separately. 
The final outputted compressed representations are in a binary string format. In order to feed these compressed representations, we need to run them through the initial layers of the decoder in order to get the latent tensor representations. 
We note this step as part of pre-processing as we are not decoding the image, we are simply obtaining the latent representation $\hat{y}$ and the associated standard deviations $\hat{\sigma}$. 
Additionally, we could not use the tensorflow-compression module out of box as we needed access to the internal operation of the decoder in-order to perform this step. As a results, we switched to a CompressAI \cite{9707070}, a Pytorch Implementation which is a partial port of the official TensorFlow compression library \cite{tfc_github}. 

Next we implemented the cResNet-39 model to solve the downstream task of texture recognition. \citet{https://doi.org/10.48550/arxiv.2104.10065} setup stated they used an ImageNet-trained cResNet-39 model from \citet{https://doi.org/10.48550/arxiv.1803.06131}. As the pretrained model was not readily available we attempted to create a similar setting to start from. Instead of training a cResNet-39 model from scratch, we stripped a pretrained ResNet-50 model of its first 11 layers. We appended the first layer with 2 separate residual blocks into which we separately feed $\hat{y}$ and $\hat{\sigma}$ after which the result is concatenated and fed forward, and finally we swapped the last Linear layer to accommodate predictions for 23 classes. We then train this model on the latent representations of the MINC-2500 dataset.

The hyper-parameters, dataset splits and experimental setup is similar to that specified in \citet{https://doi.org/10.48550/arxiv.2104.10065}. That is, we use the train-validation-test split
1 provided in the dataset, with 2125 training images, 125 validation images and 250 testing images for each class for 3 quality rate settings of 1, 4 and 8, (where 1 for lowest quality/rate and 8 for highest quality/rate). As specified in the paper, we also employed the same preprocessing to the input compressed domain representations, i.e. they are resized to 32 × 32 and then randomly cropped to 28 × 28, followed by a random horizontal flipping.

The cResNet-39 model was also incorporated to determine the perfromance of convolution based neural networks for representation of compressed images, to show it's usability for automated pipelines. 

We extended our experiments to two sets of satellite images namely RESISC-
45 and RSCNN-7 each containing different terrains classified based on satellite images. We present the results and analysis for these below.The RSCNN7 and RESISC45 datasets contain 7 scenes from satellite images each with 400 images and 45 scenes from satellite images with 700 images. Such a large dataset required a lot of time to train these networks. Similar data augmentation was applied to make the model robust to such changes in the compressed domain.

Lastly for our final set of experiments we attempt to fine-tune the encoder and downstream model at the same time. The idea behind this being that the compressed representation output from the encoder has features relevant to the downstream tasks and at the same time being relevant to the human visual system, i.e. optimizing PSNR score. We attempted to do this by minimizing the MSE Loss of the downstream tasks along with the reconstruction loss of the compression model. However since both losses are divergent in nature, we weren't able to achieve useable results. We then attempted to fine-tune the encoder on MSE Loss and achieve better results.

Our implementation code is located at \url{https://github.com/kimberley-faria/learning_based_img_compression}. The \href{https://github.com/kimberley-faria/learning_based_img_compression/blob/main/pytorch_compression-results.ipynb}{pytorch\_compression-results.ipynb} can be run to reproduce our results.

\subsection{Results and Analysis}

Table \ref{table:1} reports our results of the first baseline experiment. We report the validation accuracy for the 3 quality/rate settings and compare these to the results in \citet{https://doi.org/10.48550/arxiv.2104.10065}. A few initial observations suggest that the results are within our expectations. Firstly, the model trained on the higher quality compressed representations performs better than that of lower quality setting. We can also see the losses are higher for lower quality/rate compressed representations, evidencing the fact that the lower rate latent representation makes the downstream task harder, as less information is present for the model to use. Additionally, the cResNet-39 architecture for the 8 bpp setting has 320 channels for representation as compared to 192 channels for the 1 and 4 bpp settings, and could also account for the huge differences in performance between the 3 settings. 

We note a few implementation details that could possible contribute to the reduction in downstream performance. Firstly, the processing step (resizing followed by randomly cropping followed by randomly horizontal flipping) is done on the fly as they are being consumed by the model during training. This could result in a harder task setting as compared to prepossessing the images before beginning training. Next we note that we are not starting from an ImageNet-trained cResNet-39 model as stated above, our bootstrapped cResNet-39 initial backbone contains ImageNet-trained ResNet-50 weights.

Table \ref{table:3} summarize the results for RSCNN7 and RESISC45 datasets in the classification on compressed domain representation setting. RSCNN7 being a simpler task with just 7 classes performs well in this setting. RESISC45 on the hand has sub-par results for lower bits per pixel settings. But, it is useful to note that in the 8 bpp setting, this method achieves 81.91\% accuracy in comparison with the SOTA Resnet-50 that achieve 96.83\% accuracy. Hence it's utility in the domain is of utmost consideration. 

Table \ref{table:4} summarize the results for joint training the encoder setup and downstream model. The encoder learns the best representations to extract from the image to improve downstream model accuracy. This joint training focuses exclusively on improving the accuracy of the machine vision tasks.

\subsection{Implementation considerations}

We cover the challenges in terms of implementation:
\begin{enumerate}
    \item A significant amount of time went into understanding the learning-based compression pipeline itself. Additionally, we also needed to understand how the compressed representations fed into the downstream model and correctly specify the input layer architecture of the downstream model, as noted above we could not directly feed the string compressed representations into the downstream model.
    \item A consequence of this was that the tensorflow-compression library could not be used out-of-box, as only the metagraphs are provided for direct use and we required the pretrained models in order to perform the initial preprocessing done in the decoder. Therefore, we switched to CompressAI\cite{begaint2020compressai}, which was a partial port of the library in Pytorch. 
    \item Working on multiple datasets and multiple downstream tasks required more memory and compute resources and in conjunction with divergent lossses led to less focus on that matter. 
    
\end{enumerate}

\begin{figure}[!ht]
\centering
  \includegraphics[scale=0.6]{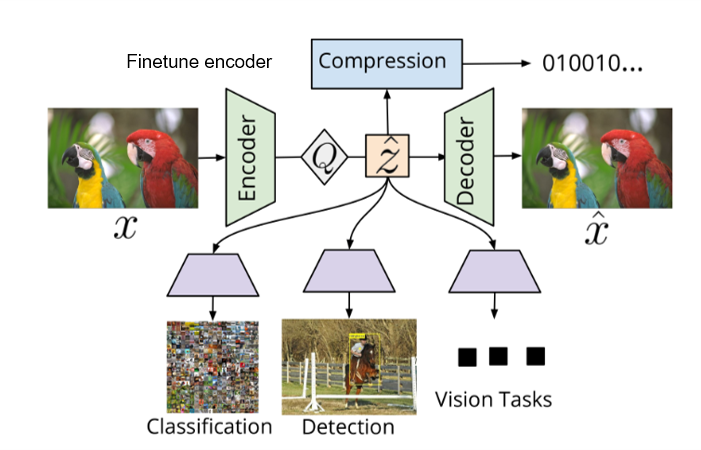}
  \caption{Utility of a trained pipeline for downstream vision tasks}
  \label{fig1.3}
\end{figure} 




\section{Conclusion}





There are several takeaways from the investigation in the project that we would like to shed a light on. This includes
\begin{itemize}
    \item Significant improvement in vision task accuracy via jointly trained model
    \item Specialized compression modules for downstream machine learning tasks help model perform better 
    \item Compression modules used without fine-tuning are also a viable option for their utility in downstream tasks without decompression, as seen for experiments on RSCNN and RESISC datasets. \cite{resisc}
\end{itemize}
Major challenges that we faced was regarding the loss functions and passing gradients from the downstream resnet model to the hyperprior encoder network. While understanding the balance between decode image quality and downstream ml task perfromance we found out that the losses are quite divergent and hence harder to train. As a result we decided to investigate the piplenine for the more important ML tasks which are key for automation pipleines that require almost no human intervention. For further investigation these are a few considerations to look forward to
\begin{itemize}
    \item Explore PSNR improvements (for human-in-loop) via training the decoder
    \item With this consideration it would be worth exploring using both the losses in tandem (human perception loss and model performance loss), although their diverging loss criteria was a challenge that we are yet to overcome.
    \item We can focus on alternate step wise training, multi-task learning approaches.
    \item Focus on multiple downstream applications for a single module to achieve a more generalized AI application focused compression module. 
\end{itemize}
\newpage
\bibliography{bibliography}
\bibliographystyle{acl_natbib}

\end{document}